# Modelling and closed loop control of near-field acoustically levitated objects

Dotan Ilssar [a], Izhak Bucher [b,*] and Henryk Flashner [c]

[a] Dynamics Laboratory, Faculty of Mechanical Engineering, Technion – Israel Institute of Technology, Haifa 3200003, Israel. dilssar@technion.ac.il

[b] Dynamics Laboratory, Faculty of Mechanical Engineering, Technion – Israel Institute of Technology, Haifa 3200003, Israel. bucher@technion.ac.il

[c] The department of Aerospace and Mechanical Engineering, University of Southern California, Los Angeles, CA 90089, USA. hflashne@usc.edu

[*] Corresponding author:

Faculty of Mechanical Engineering, Technion City, Haifa 3200003, Israel

Tel: 972-77-8873153, email address: bucher@technion.ac.il

# Abstract

The present paper introduces a novel approach for modelling the governing, slow dynamics of near-field acoustically levitated objects. This model is sufficiently simple and concise to enable designing a closed-loop controller, capable of accurate vertical positioning of a carried object. The near-field acoustic levitation phenomenon exploits the compressibility, the nonlinearity and the viscosity of the gas trapped between a rapidly oscillating surface and a freely suspended planar object, to elevate its time averaged pressure above the ambient pressure. By these means, the vertical position of loads weighing up to several kilograms can be varied between dozens and hundreds of micrometers. The simplified model developed in this paper is a second order ordinary differential equation where the height-dependent stiffness and damping terms of the gas layer are derived explicitly. This simplified model replaces a traditional model consisting of the equation of motion of the levitated object, coupled to a nonlinear partial differential equation, accounting for the behavior of the entrapped gas. Due to the relatively simple form of the model developed here, it constitutes a convenient foundation for model based control algorithms, governing the slow dynamics of near-field acoustically levitated objects. Indeed, based on the former, a height dependent, gain scheduled PID controller is developed and verified numerically and experimentally, both providing satisfying results.

# 1. Introduction

During handling and transportation of silicon wafers throughout inspection and manufacturing processes, the microelectronics industry uses conveyers, chucks and robotic arms, making mechanical contact with the substrates. Such contact generates particles, contaminating the highly controlled work environment, and thus affecting the yield significantly. To overcome this problem it was proposed to utilize the near-field acoustic levitation phenomenon, allowing a controlled levitation and transportation of the wafers without any physical contact [1]. The near-field acoustic levitation phenomenon uses high frequency, ultrasonic oscillations of a driving surface to build high-pressured layer of gas, commonly known as squeeze-film, between the driving surface and the handled object (e.g. a silicon wafer). The abovementioned pressure elevation originates from the compressibility of the entrapped gas, allowing to increase the average pressure inside the squeeze-film, above the ambient, atmospheric pressure. The pressure elevation also depends on the viscosity of the gas, preventing the latter from leaking out of the film. As a result, a load carrying force is produced, levitating the carried object above the driving surface, assuming the former is freely suspended. The ultrasonic oscillations also produce rapid pressure fluctuations, but the resulting oscillations experienced by the levitated object are considerably attenuated by the inertia, and are often in the nanometer scale.

The dynamic behavior of a near-field acoustically levitated object is commonly modelled by its equations of motion, coupled with the equation governing the flow regime inside the squeeze film (e.g. [2,3]). However, for practical applications such as controlling the dynamics of the levitated object, such model is too complex. Ilssar and Bucher [4] developed a manageable, second order ordinary differential equation, describing the more significant, slow component of the levitated object's vertical motion. This model incorporates the conservative levitation force originating from the compressibility of the gas, and the damping force acting due to its viscosity, explicitly. Yet, since the model in [4] is restricted to the case where the driving surface oscillates uniformly as a piston, the simplified model, previously suggested by Ilssar and Bucher [4] cannot describe most realistic systems. Namely, it cannot describe the dynamics of high performance systems where the driving surface exhibits spatial deformations, resulting in a non-uniform squeeze-film (e.g. [5]).

In the present paper, the aforementioned simplified model is generalized and extended for the case where the driving surface oscillates as a non-uniform standing wave. This is done following a calibration process used to adjust the height dependent levitation force to a specific form of excitation. This calibration process can be performed following either a numerical or experiment based approach. The numerical approach is based on a finite differences scheme of Reynolds equation, accounting for the flow regime inside the film (e.g. [6,7]). Whereas the experiment based approach relies on measurements, and a signal processing stage relating the levitation height at steady state to the magnitude of the excitation.

The model developed in this paper is then utilized to describe an experimental setup, enabling the formulation of a model based control loop, governing the slow dynamics of a levitated object. Indeed, after a satisfactory experimental validation of the model, the latter is exploited in order to devise a height dependent, gain-scheduled PID controller, providing rapid and accurate positioning.

## 2. Statement of the problem

Fig. 1 presents a schematic layout of the system studied in this paper. Here, a freely suspended planar object is levitated due to the elevated average pressure produced by a driving surface, oscillating at a constant frequency $\omega$ and according to a designated spatial profile $a$. From efficiency considerations, the driving surface is assumed to oscillate only at resonance, namely, its spatial displacement is determined by a single real eigenfunction [8]. Therefore, and assuming axisymmetry, the spatial profile $a$ depends merely on the radial coordinate $r$. The levitated object and the driving surface are axisymmetric, parallel, and have the same outer radius $r_0$, implying that the only DOF (degree of freedom) of the levitated object considered here, is its vertical position. Obviously, the instantaneous position of the levitated object is determined by equilibrium of its inertia, the gravity and the forces induced by the pressure exerted on it by the surrounding fluid.

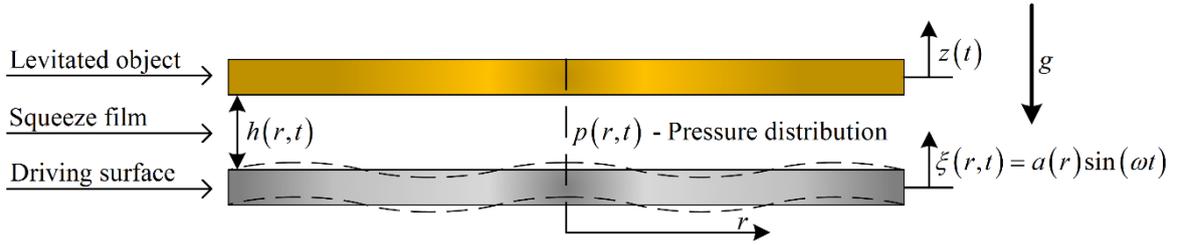

Fig. 1. Schematic layout of the axisymmetric acoustic levitation system, showing the driving lower surface and the levitated object.

As shown in Fig. 5, displaying in solid black lines the overall dynamics of the system, and by former experiments and analyses (e.g. [2,4]), the dynamics of the system illustrated in Fig. 1 can be represented as a superposition of two time scales. The first time scale relates to the excitation and the resulting low amplitude rapid oscillations experienced by the levitated object, whereas the second time scale is associated with the evolution of the levitated object, which is 2-3 orders of magnitude slower than the excitation. Consequently, the air-gap between the driving surface and the levitated object can be decomposed as follows

$$h(r,t) = z(t) - a(r)\sin(\omega t) = \bar{h}(t) + \chi(t) - a(r)\sin(\omega t). \tag{1}$$

Here, $t$ denotes the time, and $z$ is the overall dynamics of the levitated object, consisting of $\chi$ and $\bar{h}$, representing the rapid oscillations and slow evolution of the levitated object, respectively. It is important to note that $\bar{h}$ equivalently represents the slow evolution of the air-gap, since the driving surface does not experience motions relating to the slower time scale.

*2.1. The governing equations*

In order to formulate the dynamics of the system illustrated in Fig. 1 non-dimensionally, the following measures are defined:

$$P=\frac{p}{p_a}, \quad H=\frac{h}{h_0}, \quad \bar{H}=\frac{\bar{h}}{h_0}, \quad R=\frac{r}{r_0}, \quad T=\omega t, \quad \sigma=\frac{12\mu\omega r_0^2}{p_a \bar{h}^2}=\frac{12\mu\omega r_0^2}{p_a h_0^2 \bar{H}^2}, \quad \varepsilon=\frac{a}{\bar{h}}=\frac{a}{h_0 \bar{H}}, \quad (2)$$

where $p, p_a$ are the pressure distribution inside the squeeze-film, and the ambient pressure respectively, $h_0$ denotes a reference, typical clearance between the driving surface and the levitated object, and $\mu$ stands for the dynamic viscosity of the fluid. As demonstrated in the following section, the non-dimensional excitation function $\varepsilon \ll 1$, and the squeeze number $\sigma$, govern the system's instantaneous behavior (e.g. [4,7]).

The dynamics of the investigated system is represented by two equations, coupling the behavior of the fluid residing inside the squeeze film, with the dynamics of the levitated object. Under the assumptions of isothermal conditions [6,7] and axisymmetric geometry, Reynolds equation used to approximate the behavior of the entrapped gas, takes the following form [4]:

$$\frac{\partial}{\partial R}\left(RH^3 P \frac{\partial P}{\partial R}\right) = \sigma \bar{H}^2 R \frac{\partial(PH)}{\partial T}. \quad (3)$$

As demonstrated extensively in the literature (e.g. [6,7]), Reynolds equation is a simplification of the momentum and the continuity equations, assuming the fluid behaves as an ideal gas. This simplification is based on the assumption that the clearance between the levitated object and the driving surface is considerably smaller than the lateral dimensions of the system, resulting in omission of the inertia terms.

In order to consider the dynamics of the levitated object, equation (3) is coupled to its equation of motion, taking into account the body forces and the pressure exerted by the surrounding fluid. Thus, in terms of the air-gap and pressure distribution, the equation of motion of the levitated object is given by:

$$\frac{\partial^2 H}{\partial T^2} = \frac{2\pi p_a r_0^2}{m h_0 \omega^2} \int_0^1 R(P-1)\,dR - \frac{g}{h_0 \omega^2} + \varepsilon \bar{H} \sin(T) \quad (4)$$

where $g$ denotes the acceleration due to gravity.

Finally, due to the system's axisymmetry, there is no pressure gradient at the center of the film. Moreover, it is customary to assume that the pressure at the film's edge equals to the ambient pressure. This assumption was found quite accurate if both bounding surfaces have the same dimensions, but may need a correcting leakage term otherwise [9,10]. Thus, since the driving surface and the levitated object have the same radius, the boundary conditions of the pressure inside the film are taken as

$$\left.\frac{\partial P(R,T)}{\partial R}\right|_{R=0,T} = 0, \quad P(R=1,T) = 1. \tag{5}$$

## 3. A simplified model of the slow dynamics

From Fig. 5, it is clear that the displacements corresponding to the slow time scale, are considerably larger than the amplitudes of the levitated object's fast oscillations. Thus, for applicable uses such as controlling the levitation height precisely, it is usually sufficient to have a model representing merely the slow dynamics of the system, but such a model was not made available yet in the literature. Ilssar and Bucher [4] have developed a simplified model describing the slow dynamics of a near-field acoustically levitated object, using a single second order ordinary differential equation. However, this model refers to the driving surface as a rigid piston, meaning that it oscillates at a uniform amplitude. In this section, the previous simplified model is generalized in order to describe the slow dynamics of the system illustrated in Fig. 1, where the driving surface vibrates according to a more realistic, arbitrary axisymmetric profile.

*3.1. Development of a general simplified model*

At the analysis presented by Ilssar and Bucher [4], it was shown that the contribution of the surrounding fluid to the dynamic response of a near-field acoustically levitated object, can be decomposed into two distinct forces; the levitation force originating from the compressibility of the fluid, and the damping force associated with its viscosity. Moreover, it was shown that the motions related to the fast time scale could be ignored when calculating the damping force related to the slow evolution. The latter implies that the excitation form (i.e. the mode shape of the driving surface being excited) does not affect the relevant dissipative force, and so, the expression formulated for the case of a uniform excitation suits the general case presented here. Thus, the damping force is taken as following (see [4], equations 35-36):

$$F_{damping} = -\frac{3\pi\mu\omega r_0^2}{2 p_a h_0^2 \bar{H}^3} \frac{d\bar{H}}{dT}, \tag{6}$$

where $F_{damping}$ is normalized by $p_a r_0^2$.

To complete the formulation of the generalized simplified model, all that left is to recalculate the slowly varying term of the levitation force. This force results from the pressure built by the oscillations of both the driving surface and the levitated object, due to the compressibility of the gas. Thus, since the evolution of the system is much slower than the fast oscillations, it hardly affects the compression of the gas residing inside the squeeze film, meaning that its contribution to the levitation force can be neglected. Moreover, it was already shown (e.g. [2–4]) that the amplitudes of the levitated object's fast oscillations are 2-3 orders of magnitude smaller than the excitation magnitude and so their contribution to the levitation force can be neglected as well. Namely, the levitation force is hardly affected by both slow and fast dynamics of the levitated object, and so, its slowly varying term can be approximated as a static force acting on a fixed object placed at a constant nominal clearance $\bar{H}$ from the driving surface.

To approximate the time averaged levitation force acting on a fixed object using an asymptotic analysis, a small parameter needs to be defined. As mentioned above, the magnitude of the non-dimensional excitation function $\varepsilon$ is small; however, since this measure depends on the spatial coordinate $R$, it cannot function as the small parameter in a regular perturbation type of analysis. Therefore, the small parameter is chosen as the height dependent norm of $\varepsilon$, separating the non-dimensional excitation function as follows:

$$\varepsilon(\bar{H},R) = \|\varepsilon(\bar{H},R)\| \hat{\varepsilon}(R) = \delta(\bar{H}) \hat{\varepsilon}(R). \tag{7}$$

Here, since the parameter $\delta$ takes the small magnitude of the non-dimensional excitation function, the norm of $\hat{\varepsilon}$ equals to unity. Moreover, for sake of consistency with the model describing the behavior of the system under uniform excitation [4], the weight function of the norm is chosen as 1.

According to (7), when the nominal air-gap is constant and equals to $\bar{H}$, the air gap varies as following:

$$H(T) = \bar{H}\left[1 - \delta(\bar{H}) \hat{\varepsilon}(R) \sin(T)\right]. \tag{8}$$

Additionally, in order to take nonlinear effects into account, the pressure distribution is represented as a second order asymptotic series, as suggested by Minikes *et al.* [11]:

$$P(R,T) = 1 + \delta \Pi_A(R,T) + \delta^2 \Pi_B(R,T) + \mathcal{O}(\delta^3). \tag{9}$$

Substituting (8) and (9) into Reynolds equation (3) results in an expression consisting of several orders of $\delta$. In order to balance the leading order equation associated with the terms of order $\delta$, the following harmonic solution is suggested:

$$\Pi_A(R,T) = \Pi_1(R)\cos(T) + \Pi_2(R)\sin(T). \tag{10}$$

Harmonic balance of the leading order equation using (10), yields two coupled ordinary differential equations from which the spatial functions $\Pi_1, \Pi_2$ can be determined, assuming the excitation function is known.

As shown in the past (e.g. [4,11]), substitution of (10) into the second order equation, comprising the terms of order $\delta^2$, gives rise to terms dependent on second temporal harmonics and to time independent terms. Therefore, in order to balance this equation, the following solution is proposed:

$$\Pi_B(R,T) = \Pi_3(R)\cos(2T) + \Pi_4(R)\sin(2T) + \Pi_5(R). \tag{11}$$

Substituting (10) and (11) into the second order equation, and averaging over one excitation period provides an ordinary differential equation, depending on $\Pi_1, \Pi_2, \Pi_5$ and $\hat{\varepsilon}$. Obviously, assuming $\Pi_1, \Pi_2, \hat{\varepsilon}$ and are known, the function $\Pi_5$ representing the time independent part of the gauge pressure distribution, can be obtained.

Since the pressure fluctuations are irrelevant when calculating the time averaged levitation force, than if $\Pi_5$ is known, a second order approximation of this force can be found. This is done by integrating the total pressure acting on the levitated object from both its sides, over its area, and averaging over one excitation period, as shown below:

$$F_{levitation} = \frac{1}{2\pi}\int_0^{2\pi}\int_0^{2\pi}\int_0^1 R(P-1)\mathrm{d}R\mathrm{d}\theta\mathrm{d}T = 2\pi\delta^2\int_0^1 R\Pi_5(R)\mathrm{d}R = \pi\delta^2 K(\sigma), \qquad (12)$$

where $F_{levitation}$ is normalized by $p_a r_0^2$.

The function $K(\sigma)$ is a calibration function of a particular set-up and it contains information about the spatial pressure distribution that depends on the excitation form and on the squeeze number. This function captures the relatively complex features of the levitation process, considering only the slow dynamics (quasi-static compared with the excitation frequency).

It should be noted that a closed form of the spatial function $\Pi_5$ is usually hard to find. Thus typically, an explicit expression of the function $K(\sigma)$, relating the time averaged levitation force and the squeeze number, cannot be formulated. However, $K(\sigma)$ can be computed numerically or experimentally, as shown in the following sub-section. The latter leads to a closed expression for the time averaged levitation force.

The influence of the total pressure acting on the levitated object was decomposed into two forces whose slowly varying terms are denoted as $F_{levitation}$ and $F_{damping}$. A superposition of these terms constitutes a second order approximation of the pressure term in the equation of motion (4), averaged over one excitation period. Consequently, averaging (4) over one excitation period and using (6),(12), yields the following equation, describing the slow dynamics of the system illustrated in Fig. 1:

$$\frac{\mathrm{d}^2\bar{H}}{\mathrm{d}T^2} = \frac{C_s^* u^2 K(\sigma(\bar{H}))}{\bar{H}^2} - \frac{C_d}{\bar{H}^3}\frac{\mathrm{d}\bar{H}}{\mathrm{d}T} - G. \qquad (13)$$

Here, $u$ representing the excitation magnitude, $C_s^*$ denoting a coefficient of the stiffness term, $C_d$ standing for the coefficient of the damping term and $G$ symbolizing the gravity, are given as followings:

$$u = \|a(r)\|/h_0, \quad C_s^* = \pi r_0^2 p_a/m\omega^2 h_0, \quad C_d = 3\pi\mu r_0^4/2m\omega h_0^3, \quad G = g/\omega^2 h_0. \qquad (14)$$

Obviously, here the levitated object is no longer fixed, and therefore the squeeze number $\sigma$ is a functional of the slow evolution $\bar{H}$.

*3.2. Finding an approximation for the calibration function*

As stated above, the slow dynamics of a near-field acoustically levitated object can be formulated by (13)-(14). However, finding a closed form for the function $K(\sigma)$ is not an easy task. Therefore, in this sub-section, two alternative approaches, leading to an approximation of this function are presented. The first approach is based on a numerical scheme, describing the behavior of the squeeze-film according to Reynolds equation (3), whereas the second approach is experimental, and based on the steady state equation obtained from (13) by nulling the time derivatives of $\bar{H}$.

*3.2.1. The numerical approach*

Since the motions of the levitated object were neglected in the development of the slowly varying levitation force, the function $K(\sigma)$ should be calculated neglecting these motions as well. Indeed, the first approach to approximate $K(\sigma)$ is based on a numerical scheme describing the degenerated form of the system illustrated in Fig. 1, where the upper object is fixed. The abovementioned scheme consists of numerous ordinary differential equations achieved by discretizing Reynolds equation (3) as following:

$$\left(\frac{\partial P}{\partial T}\right)_n = \frac{1}{\sigma \bar{H}^2}\left[\frac{H_n^2 P_n}{R_n}\left(\frac{\partial P}{\partial R}\right)_n + 3H_n P_n \left(\frac{\partial H}{\partial R}\right)_n\left(\frac{\partial P}{\partial R}\right)_n\right] \\ + \frac{1}{\sigma \bar{H}^2}\left[H_n^2\left(\frac{\partial P}{\partial R}\right)_n^2 + H_n^2 P_n\left(\frac{\partial^2 P}{\partial R^2}\right)_n\right] - \frac{P_n}{H_n}\left(\frac{\partial H}{\partial T}\right)_n \quad (15)$$

where $n \in [1, N]$ denotes the index of the radial node residing at $R = (n-1)/(N-1)$.

In each of these equations, the different spatial derivatives are calculated using central differences formulas, while the needed values at the external nodes $n = 1, N$ are determined according to the boundary conditions (5). Furthermore, assuming the excitation function is given, the air-gap $H$ at each node is known in every instance, leaving the pressure as the only unknown function, whose instantaneous values are calculated from (15) using numerical integration.

It should be noted that the derivation presented in the previous sub-section neglects the transient behavior of the pressure. The latter is a valid assumption since the time constants of the pressure, are considerably lower than those of the levitated object's evolution. Therefore and according to (12), the time averaged levitation force is calculated by averaging the pressure obtained from (15), over an integer number of excitation periods and integrating over the area, after convergence to steady state.

According to (12), calculation of $F_{levitation}(\sigma)$ for various squeeze numbers in the desired work range, and division by $\pi \delta^2$ results in the desired values of $K(\sigma)$. Here, in order to satisfy the asymptotic assumptions led to the simplified model (13), implying that the contribution of $\delta, \sigma$ to the levitation force can be separated, $\delta$ should be small in the simulations as well. Finally, in order to use

$K(\sigma)$ in the simplified model (13), a closed form analytical expression needs to be found. This is done using curve fitting on the values calculated numerically.

*3.2.2. The experiment-based approach*

The second approach for approximating the function $K(\sigma)$ is empirical and requires measuring the time averaged levitation heights at steady state $\bar{H}_{eq}$, for various constant excitation magnitudes $u_{eq}$. Substitution of $\bar{H}_{eq}, u_{eq}$ into the simplified model (13) eliminates the time derivatives of $\bar{H}$, resulting in the following equation:

$$C_s^* u_{eq}^2 K\left(\sigma\left(\bar{H}_{eq}\right)\right) = G\bar{H}_{eq}^2. \tag{16}$$

The transcendental equation (16) obtained for every pair $\bar{H}_{eq}, u_{eq}$, provides a single value of $K(\sigma)$. Curve fitting on the values achieved from the different equilibrium points, yields a continuous approximation that can be used in (13).

Obviously, the experiment-based approach is more reliable than the numerical approach since it depends on experimental data, and thus helps correcting inaccuracies in the model and including overlooked effects. The inaccuracies originate from restricting assumptions such as pressure release boundary conditions, stating that the pressure at the peripheries of the film equals to the ambient pressure. Furthermore, is should be noted that the experimental-based approach can consider additional static forces, such as magnetic forces acting on the system (see sub-section 3.3). Namely, such forces are considered as supplementary terms on the right hand side of (16), thus this approach takes these terms into account when calculating $K(\sigma)$.

*3.3. Simplified modelling of the experimental setup*

The experimental work presented in this paper was performed using the setup displayed in Fig. 2. This setup consists of a levitated object whose weight and radius are 134.7 g and 50 mm respectively, and a custom made piezoelectric actuator supplying the required excitation. The actuator is an assembly of a Langevin transducer, and a structure used to magnify the displacements at its base, for production of significant excitation amplitudes. From efficiency considerations both components comprising the actuator are designated to operate at approximately 28.5 kHz. At this frequency, all axial elements resonate at their first elastic longitudinal mode, and the upper plate whose top surface functions as the driving surface, resonates at its second axisymmetric flexural mode (see Fig. 3). In order to eliminate undesired degrees of freedom such as lateral and tilting oscillations [12], a magnetic centering array was implemented. As shown in Fig. 2, this array uses magnetic repulsion forces between three centering elements and a series of magnets placed on the perimeter of the levitated object.

It should be noted that in order to preserve high performance, and to comply with the model's assumption, referring to the excitation form as a pure standing wave, in all of the experiments discussed in this paper, the excitation frequency was determined according to a digitally controlled Phase-Locked loop algorithm (e.g. [13]). However, since in all of the experiments the frequency changed by less than 0.1%, these variations are not taken into account in the analyses.

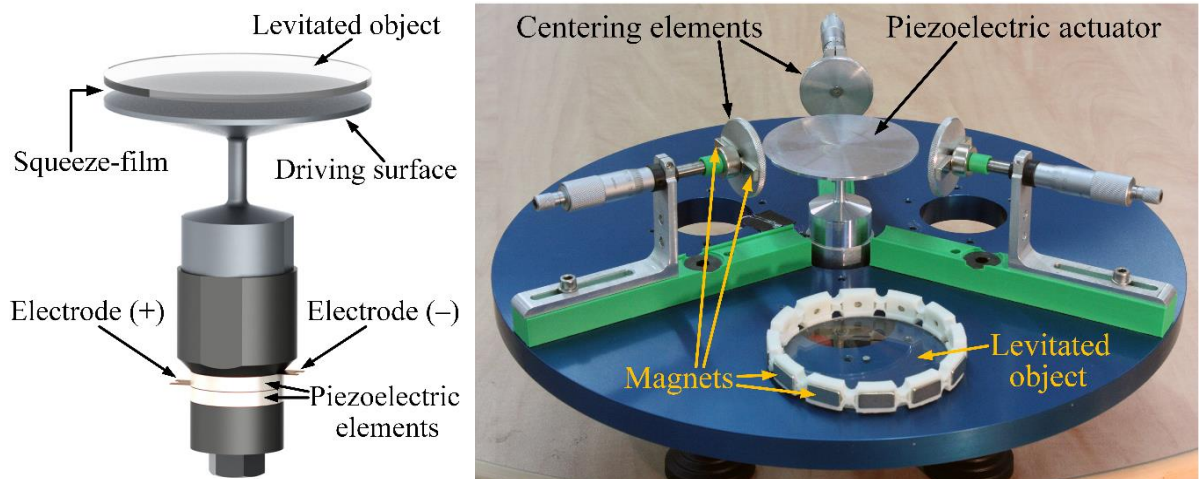

Fig. 2. Schematic layout and a photograph of the experimental setup. Left: The piezoelectric actuator and a sample levitated object. Right: The experimental rig showing the piezoelectric actuator in the middle and the levitated object with centering passive magnets (set aside for visibility).

To model the experimental system according to (13), and to evaluate the agreement between the two approaches suggested for approximating $K(\sigma)$, three sets of experiments were carried out. In the first experiment-set, the excitation form $\hat{\varepsilon}$, which is the mode-shape of the driving surface being excited, was measured using a laser Doppler sensor (Polytec™, OFV-303). These measurements were performed under numerous different excitation magnitudes in the desired working range, when the system was loaded with the mass mentioned above. According to these experiments, the excitation form does not change with the levitation height, thus as illustrated in Fig. 3, $\hat{\varepsilon}$ is taken as a levitation-height independent function, complying with (7).

The purpose of the second experiment-set was to find the relation between the excitation magnitude $u$, and the amplitude of the voltage supplied to the system, denoted as $V$. For this sake the relation between $V$ and the amplitude at the center of the loaded driving surface was measured using a laser Doppler sensor (Polytec™, OFV-551), under excitation with several magnitudes in the allowable range. This relation, together with the data from the first experiment-set, relating the amplitude at the center of the driving surface to $\|a\|$, led to the desired relation.

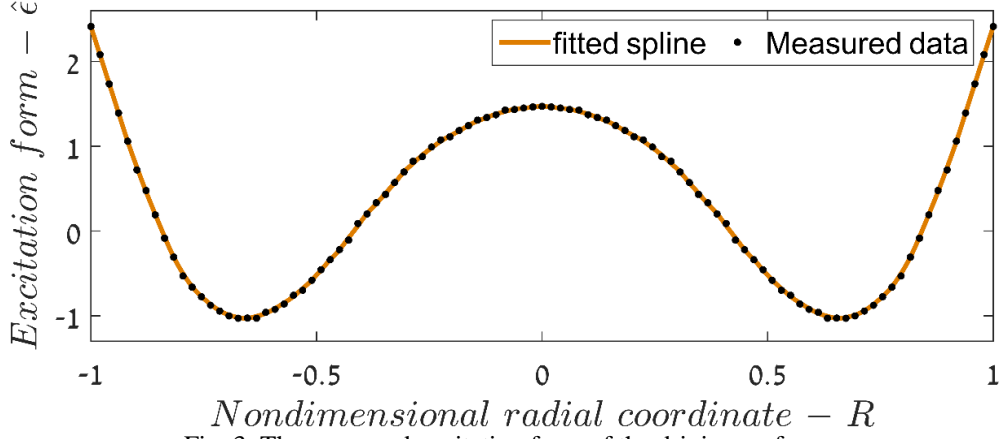
Fig. 3. The measured excitation form of the driving surface.

In the third experiment-set conducted for the identification of the system, the steady-state height of the levitated mass was measured several times under numerous excitation magnitudes, using a laser triangulation sensor (Keyence[TM], LK-H008). The latter relates the steady-state levitation height to the amplitude of the input voltage.

As explained above, knowing the spatial excitation form is sufficient for calculating $K(\sigma)$ according to the numerical approach. However, to estimate this function experimentally, the data acquired in all three experiment-sets is necessary. Namely, in the third experiment-set, the relation between the levitation height and the input voltage was found; yet, to utilize (16), the former needs to be associated with $u_{eq}$, depending on both the shape and the magnitude of the excitation function. This relation can be identified from the first and second experiment-sets, providing the mode shape and the relation between the input voltage and the excitation magnitude. Thus given the data from all three experiment-sets, the experimental estimation of $K(\sigma)$ can be found.

According to the algorithms discussed above, and using the data from all three experiment-sets, Fig. 4, comparing the numerically obtained and the experimental forms of $K(\sigma)$, was produced. This figure shows that the experimental values of $K(\sigma)$, obtained over 20 experiments are quite repetitive, and their fitted curve given by

$$K(\sigma) \approx 1.2348 + 2.2928 \cdot 10^{-2}\sigma - 1.301 \cdot 10^{-4}\sigma^2 + 2.9826 \cdot 10^{-7}\sigma^3 - 2.4985 \cdot 10^{-10}\sigma^4 \qquad (17)$$

constitutes a good approximation to these values. Thus (17) can be used as the calibration function in (13). However, one should notice a considerable difference between the values obtained from the different approaches, implying an error in the numerically obtained values of $K(\sigma)$. A significant part of this error can be attributed to the fact that the numerical approach does not consider the force induced by the magnetic array presented in Fig. 2. The axial component of the magnetic force, whose positive direction is chosen downwards, adds an additional term to (13), denoted as $F_{mag}$. The magnitude of this term is the axial projection of the magnetic force, normalized by $m\omega^2 h_0$. Therefore, when calculating $K(\sigma)$ according to the experiment-based approach, the axial component of the magnetic force, is taken into account. Namely, by substituting $\bar{H}_{eq}, u_{eq}$ into (13), and eliminating all the time derivatives, (16) takes the following form, considering the magnetic force:

$$C_s^* u_{eq}^2 K\left(\sigma\left(\bar{H}_{eq}\right)\right) = \bar{H}_{eq}^2 \left(G + F_{mag}\right) \quad (18)$$

Since the distance between the static magnets and the magnets located on the levitated object, is large compared to their depth and width, they can be treated as magnetic dipoles. Moreover, the length of the magnets and the radii of the circles on which they are located, are larger than the distance between them. Thus the axial magnetic force can be calculated referring to the magnets as three pairs of long permanent magnets. In this case the axial component of the magnetic force can be approximated proportional to $\sin(3\theta)$, when $\theta$ is the angle between the horizon and the line connecting each pair of the repulsing magnets [14]. As can be seen from Fig. 4, the deviation between the numerically obtained and the experimental values of $K(\sigma)$ is monotonous in most of the work range, with exception of small squeeze numbers, corresponding to significant air-gaps. This trend is consistent with Yonnet [14], assuming the levitated object is located below the magnetic array such that $\theta > \pi/6$ until $\sigma \approx 60$ where the slope of the deviation between the experimental and the numerically obtained values of $K(\sigma)$ changes its sign. Furthermore, one should notice that at high squeeze numbers, corresponding to low altitudes, the deviation between the experimental and the numerically obtained values of $K(\sigma)$ decreases, since the magnetic force decays.

Finally, it should be noted that at small squeeze numbers, the numerically obtained values of $K(\sigma)$ become negative, indicating that under appropriate conditions the levitation force turn into an attractive force. The latter is beyond the scope of this paper and therefore it will not be discussed here.

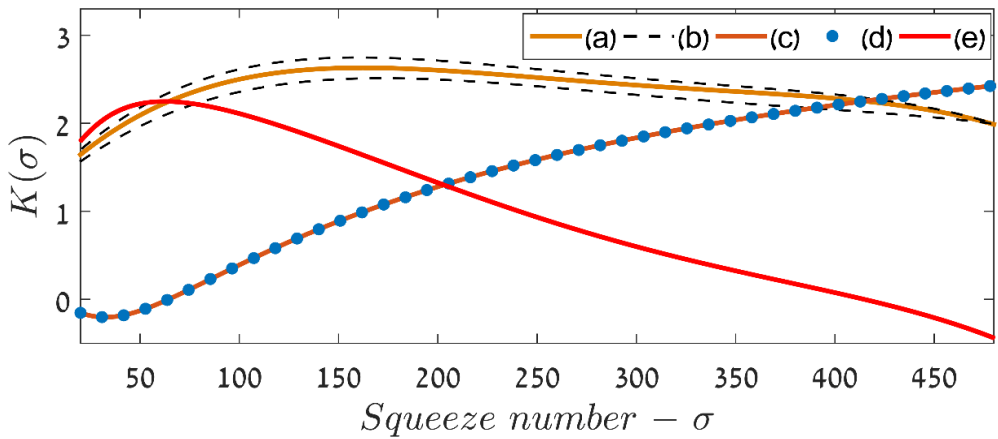

Fig. 4. Comparison between the numerically obtained and the measured (experimental) values of K(σ). (a) The curve fitted to the measurements, (b) The area enclosing all of the measurements, (c) The curve fitted to the numerically obtained values (d) The values calculated according to the numerical approach, (e) The difference between fitted curves.

It was witnessed above that there is a significant deviation in the values of $K(\sigma)$ achieved using the numerical approach, originating vastly from the fact that the forces induced by the magnetic array were not taken into account. However, it should be noted that when the excitation magnitude is much smaller than the air-gap, the simplified model (13) together with $K(\sigma)$ achieved from the numerical approach, predicts accurately the slow dynamics of the original model (3)-(5). Fig. 5 shows a comparison between number of dynamic responses whose properties are given in Table 1, achieved under the excitation form illustrated in Fig. 3, in absence of magnetic forces. These responses were obtained from the original model (3)-(5), and from the simplified model (13) with $K(\sigma)$ achieved using the numerical approach. The former was calculated numerically by solving the equation of motion (4) together with the ordinary differential equation system (15), using finite differences in space and numerical integration in time. As can be seen from Fig. 5 and Table 1, the simplified model calculated utilizing the numerical approach capture accurately the slow dynamics of the system modelled by (3)-(5), implying that this approach indeed manages to describe accurately the conservative force of this system. However, since the simplified model was developed assuming $\delta \ll 1$, as this measure grows, the agreement between the original model and the simplified model deteriorates, at steady-state and also during the transient response. It is clear from Fig. 5 that the transient response of the simplified model is more sensitive to high values of $\delta$ than its steady-state behavior.

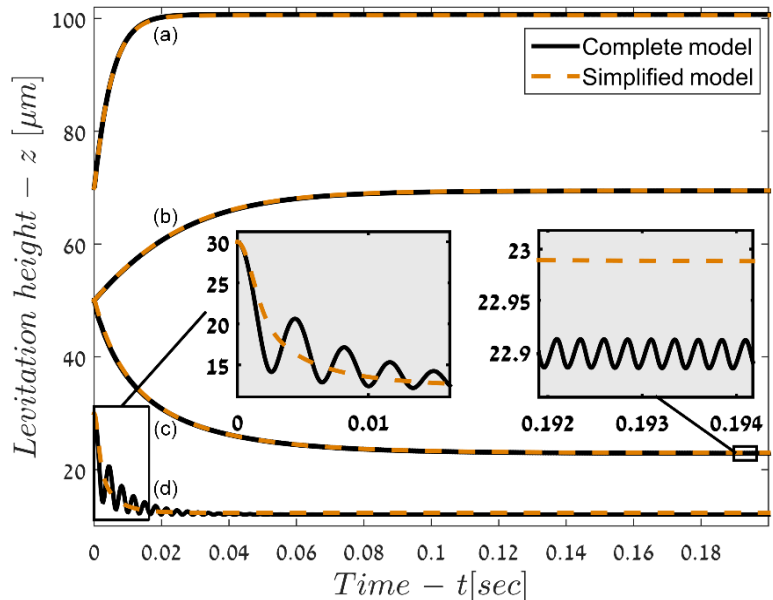

Fig. 5. Comparison between the simulated dynamic responses whose properties are given in Table 1. The solid black curves were obtained from the original (Reynolds equation based) model, and the dashed orange curves were achieved using the simplified model, utilizing the numerically obtained values of K($\sigma$).

Table 1.
Significant characteristics of the dynamic responses in Fig. 5

| Curve | Mass [g] | Excitation frequency [kHz] | ‖a‖ [μm] | δ at steady state | Steady-state error |
|---|---|---|---|---|---|
| a | 134.7 | 28.5 | 7 | 0.0696 | 0.1659% |
| b | 134.7 | 28.5 | 2.5 | 0.0360 | 0.0779% |
| c | 1000 | 4 | 2 | 0.0870 | 0.3976% |
| d | 20000 | 2 | 4 | 0.3244 | 2.3425% |

Fig. 6 presents a typical dynamic response of the levitated object, obtained experimentally under excitation with several constant magnitudes. It also displays the theoretical prediction of the response, achieved using the simplified model given by (13) and (17). This figure reveals that the deviations between the experimental and the predicted steady-state heights are less than 2%, over the entire work range. Moreover, at low altitudes the simplified model accurately predicts the dynamic behavior of the system, meaning that in addition to the stiffness term found empirically, the damping term calculated theoretically almost flawlessly describes the actual damping. However, at significant altitudes the theoretically computed damping term underestimates the real damping. For example, when the levitation height is around 190 µm, the predicted rising time is about 9 times shorter than its experimental value. One possible reason for these deviations is the fact that at significant clearances the assumptions made in the development of Reynolds equation weaken. The latter leads to loss of accuracy since the simplified model (13) is fully based on Reynolds equation.

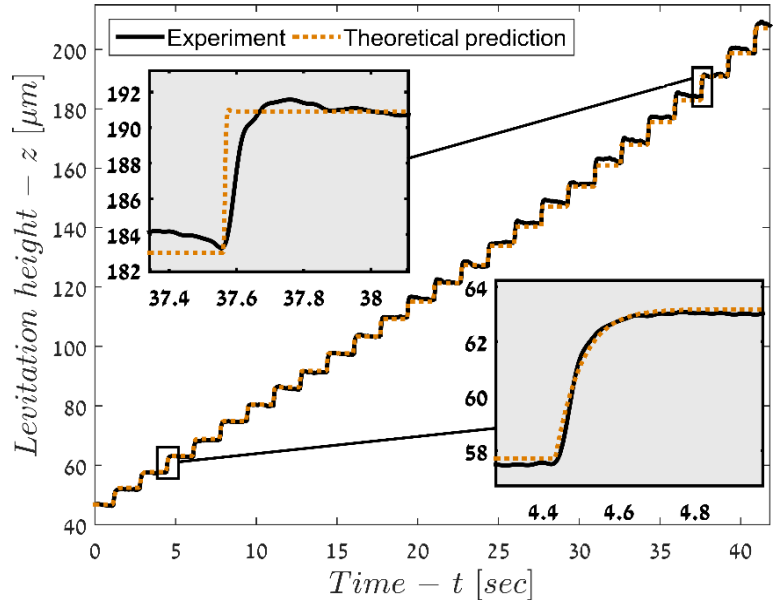

Fig. 6. The dynamic response of the levitated object under excitation with several constant magnitudes – comparison between experimental and theoretical results. The dashed orange curve is based on (13),(17) and the solid black curve is measured on the system shown in Fig. 2.

The theoretical damping term can be corrected empirically as done by Ilssar and Bucher [4]. However, since the behavior of the system is affected by temperature and moisture deviations, it seems impractical to include such lengthy procedure in the calibration process. Nevertheless, the simplified model (13),(17) seems adequate for purposes of designing a controller for the system presented in Fig. 2, assuming the controller is robust. Thus in the next section, the model (13),(17) is utilized as a basis for a control law, governing the slow dynamics of the experimental setup.

## 4. Closed loop control

Both the levitation force and the damping force comprised in the simplified model (13), were developed assuming the excitation magnitude and frequency are constant, considering terms up to order $\delta^2$. An implicit assumption of this development is that the slow dynamics of the system is sufficiently slow such that

$$\mathrm{d}\delta/\mathrm{d}T \ll \delta^2, \qquad (19)$$

implying that (13) also holds if the excitation magnitude varies slowly so the latter is still met.

The objective of the current section is to develop a control law, governing the slow dynamics of the system. Hence slow variations of the excitation magnitude, meeting the abovementioned condition, are sufficient. In this section, a gain-scheduled controller, governing the slow dynamics of the experimental setup is designed based on the simplified model (13). Next, after numerical simulations used to assess the performance of the controller, the latter is evaluated experimentally.

### 4.1. Design of a gain-scheduled controller

In the current sub-section, a gain-scheduled controller, governing the slow dynamics of the experimental setup is developed, relying on the simplified model (13),(17). Namely, the control law is based on an LPV (linear parameter-varying) model, obtained by linearizing (13) around an arbitrary equilibrium point in the desired work range, using first order Taylor expansion. The latter provides the following family of linear systems parameterized by the equilibrium point $u_{eq}, \bar{H}_{eq}$

$$\frac{\mathrm{d}^2 \Delta \bar{H}}{\mathrm{d}T^2} \approx \underbrace{\left( \frac{1}{\bar{H}_{eq}^2} \frac{\mathrm{d}K(\sigma(\bar{H}))}{\mathrm{d}\bar{H}} \bigg|_{\bar{H}_{eq}} - \frac{2K(\sigma(\bar{H}_{eq}))}{\bar{H}_{eq}^3} \right) C_s^* u_{eq}^2 \Delta \bar{H}}_{-\alpha_0} - \underbrace{\frac{C_d}{\bar{H}_{eq}^3} \frac{\mathrm{d}\Delta \bar{H}}{\mathrm{d}T}}_{\alpha_1} + \underbrace{\frac{2C_s^* u_{eq} K(\sigma(\bar{H}_{eq}))}{\bar{H}_{eq}^2}}_{\beta_0} \Delta u. \qquad (20)$$

Here, $\Delta u, \Delta \bar{H} \ll 1$ are small deviations around the equilibrium point $u_{eq}, \bar{H}_{eq}$, denoting a constant excitation magnitude and its resulting steady state levitation height, respectively. The different equilibrium points are obtained from the steady-state equation (16).

In order to achieve good performance and to eliminate steady state errors, the chosen controller is of a PID form where the proportional and the integral actions act on the error between the reference $r$ and the levitation height $\bar{H}$, whereas the derivative action acts on $\bar{H}$ [15]. Here, the three controller parameters $k_p, k_i, k_d$, denoting the proportional, integral, and derivative gains respectively, vary with the levitation height $\bar{H}$, implying that $\bar{H}$ is the scheduling variable. For sake of causality, and to avoid amplification of high frequency noise, the differentiator is connected in series to a low-pass filter whose cut-off frequency is denoted $\omega_{LPF}$, as illustrated in Fig. 7. According to the above, the gain-scheduled controller is given by

$$u = k_p(\bar{H})\left[r - \bar{H}\right] + \eta - k_d(\bar{H})\psi \tag{21}$$

where

$$\frac{d\eta}{dT} = k_i(\bar{H})\left[r - \bar{H}\right], \quad \frac{d\psi}{dT} + \omega_{LPF}\psi = \omega_{LPF}\frac{d\bar{H}}{dT}. \tag{22}$$

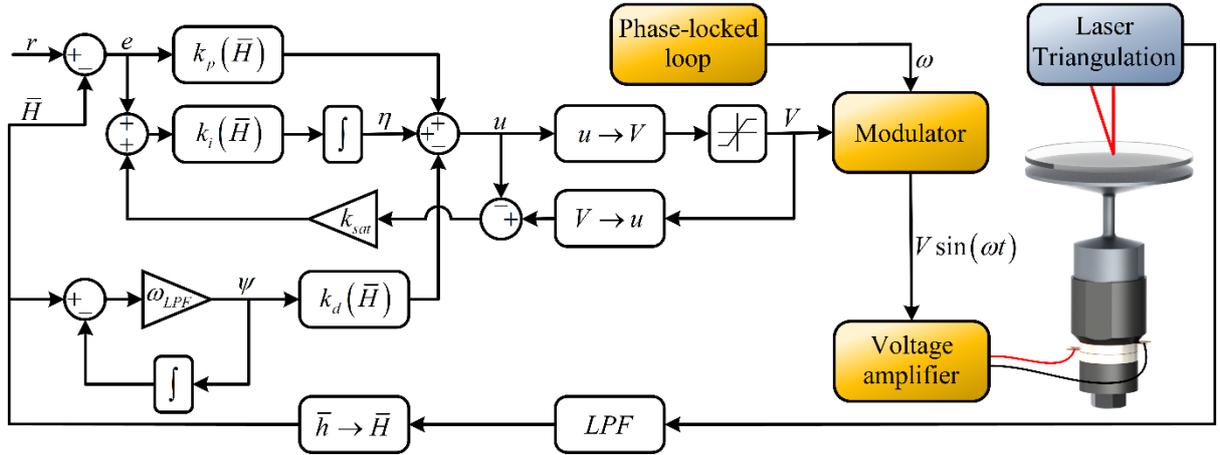

Fig. 7. Schematic layout of the closed loop system, comprising the PID gain-scheduled controller, the resonance tracking implementation and the acoustic levitation system.

The controller given by (21) and (22) is based on the velocity algorithm suggested by Kaminer *et al.* [16]. It is easy to verify that under the assumption of zero steady-state error, this controller has no hidden coupling terms. Namely, around each equilibrium point, the gain-scheduled controller is similar to the linear controller designed to achieve the desired performance at this point [17]. Thus, omitting the arguments $\bar{H}_{eq}, u_{eq}$ from the height dependent parameters $k_p, k_i, k_d, \alpha_0, \alpha_1, \beta_0$ for brevity, each local closed loop transfer function is of the following form:

$$\frac{\Delta \bar{H}}{\Delta r} = \beta_0 \frac{k_p s^2 + (\omega_{LPF} k_p + k_i)s + \omega_{LPF} k_i}{s^4 + C_3 s^3 + C_2 s^2 + C_1 s + C_0} \tag{23}$$

where

$$\begin{aligned} C_0 &= \beta_0 \omega_{LPF} k_i, \quad C_1 = \alpha_0 \omega_{LPF} + \beta_0 \omega_{LPF} k_p + \beta_0 k_i \\ C_2 &= \alpha_1 \omega_{LPF} + \alpha_0 + \beta_0 k_p + \beta_0 \omega_{LPF} k_d, \quad C_3 = \alpha_1 + \omega_{LPF}. \end{aligned} \tag{24}$$

Since the gain-scheduled controller (21),(22) equals to its corresponding linear form at each design point, the controller coefficients at every point should be chosen in a way that satisfies the required performance locally. Here, to achieve rapid convergence to steady state with reduced oscillations, the controller is designed such that around each design point, every two poles are identical and thus real valued, implying on critical damping. The abovementioned requirement reduces the controller design into two degrees of freedom, determined by setting the cut-off frequencies of the closed loop and the differentiator's low-pass filter to desired values. These frequencies are chosen to attenuate high frequency noise without affecting the system performance. It should be noted that as presented in Fig. 7, the output signal goes through an analog low-pass filter. However, the latter is not taken into account in the controller design since its cut-off frequency is chosen to be much higher than those of the closed loop and the filter connected to the differentiator. Another component that is disregarded in the controller design is an anti-windup loop, also appear in Fig. 7.

The levitation height $\bar{H}$, serving as the scheduling variable, is necessarily continuously differentiable over the entire work range due to the inertia of the levitated object. As a result the time derivative of $\bar{H}$ is bounded for every $T \geq 0$. Moreover, it should be noted that the calibration function (17) is continuously differentiable with respect to $\bar{H}$ over the entire work range, thus so as the parameters $\alpha_0, \alpha_1, \beta_0$. Furthermore, assuming the controller is designed properly, its parameters $k_p, k_i, k_d$ are taken as continuously differentiable and the closed loop is considered locally stable around every equilibrium point $u_{eq}, \bar{H}_{eq}$. Therefore, denoting the state vector of the closed loop when the scheduling variable is equal to $\bar{H}_{eq}$ as $\Upsilon(\bar{H}_{eq}) = \{\bar{H} \quad d\bar{H}/dT \quad \eta \quad \psi\}^T$, it can be shown that if the measures

$$\left\| \Upsilon_0\left(\bar{H}(T=0)\right) - \Upsilon_{ss}\left(\bar{H}(T=0)\right) \right\|, \quad \left\| d\bar{H}/dT \right\| \tag{25}$$

are sufficiently small for every $T \geq 0$, then $\Upsilon$ is uniformly bounded for all $T \geq 0$, and the error $e = r - \bar{H}$ becomes of the same order as $\left\| d\bar{H}/dT \right\|$, after a finite time. Moreover, if the scheduling variable $\bar{H}$ converges to a constant steady state value, implying that its time derivative converges to zero as $T \to \infty$, than the error vanishes as $T \to \infty$. The abovementioned properties can be proven following Khalil [18] (Theorem 12.1), who verified these properties for the velocity algorithm given in [16].

Using the parameters given in Table 2, the diagram presented in Fig. 8 was produced. This diagram shows the absolute value of the closed loop pole, closest to the stability borderline (the unit circle), after transformation to the Z domain according to forward Euler method. The latter is presented for every combination of operating point, denoting the actual levitation height, and design point, representing the height where the local controller is designed, in the work range. This diagram is presented in discrete terms because both the theoretical and the experimental controllers discussed here are implemented digitally.

Fig. 8 shows that with the chosen controller, the closed loop is locally stable for every combination of design point and operating point. Moreover, it can be shown that the controller parameters $k_p, k_i, k_d$ are continuously differentiable with respect to $\bar{H}$ in the entire work range. Therefore, according to the stability property discussed above, if the initial conditions are adequately close to their steady-state values, implying that the measures (25) are sufficiently small for all $T \geq 0$, than $\bar{H}$ is uniformly bounded, and its steady-state error to a step input converges to zero. Furthermore, it is clear that the largest stability margin is achieved when the designed point equals to the operating point, namely, when using a gain-scheduled controller.

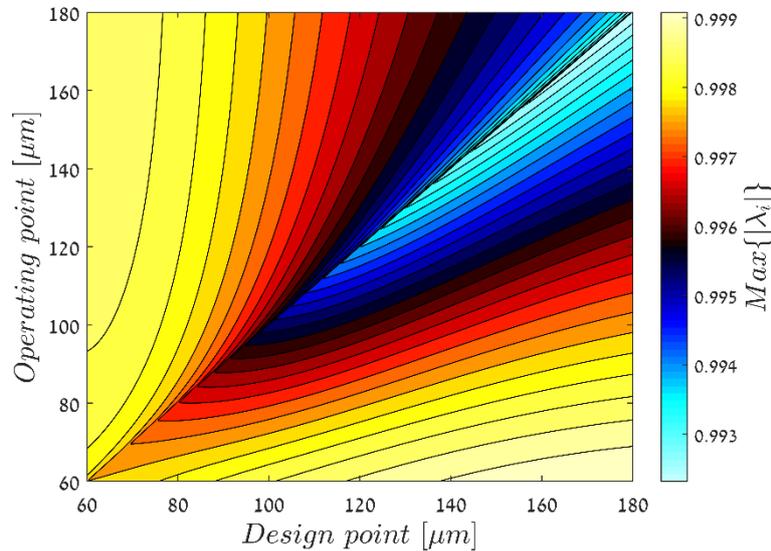

Fig. 8. The absolute value of the closed loop discrete pole, closest to the stability borderline, as a function of the designated levitation height and the actual height.

Table 2.
The parameters of the closed loop system considered in section 4.

| | |
|---|---:|
| Cut-off frequency of the closed loop | 10 Hz |
| Cut-off frequency of the differentiator's filter ($\omega_{LPF}$) | 300 Hz |
| Cut-off frequency of the analog Low-pass filter | 2000 Hz |
| Sampling rate | 2000 Hz |

As a final step for assessing the chosen gain-scheduled controller, its performance was examined utilizing a numerical scheme based on the simplified model (13),(17), describing the system illustrated in Fig. 7. Using this numerical scheme, the step response of the closed loop system to a reference signal, varying between 60 µm and 180 µm, was simulated. Moreover, as presented in Fig. 9 and Fig. 10, similar simulations, where the gain-scheduled controller was replaced with different linear controllers designed similarly for a single design point, were also performed. In order to describe a more realistic situation, measurement noise was added to all of the abovementioned simulations. However, it should be emphasized that the simplified model (13) is valid only if the condition (19) is met. To satisfy this condition, the measurement noise was taken as colored noise with maximal magnitude of 2 µm, whose cut-off frequency is 300 Hz.

Fig. 9 and Fig. 10 show the simulated dynamic responses obtained using the gain-scheduled controller and three different linear controllers. These figures demonstrate that theoretically, the former indeed provides the best performance. Namely, the gain-scheduled controller provides both noise attenuation and the fastest and least oscillatory convergence to steady state. Moreover, one should notice that there is a trade-off between noise reduction obtained using linear controllers designed for high altitudes, and good transient behavior achieved using linear controllers designed for low altitudes. To achieve both noise reduction and good transient behavior in the entire work range, the gain-scheduled controller seems as the preferable choice.

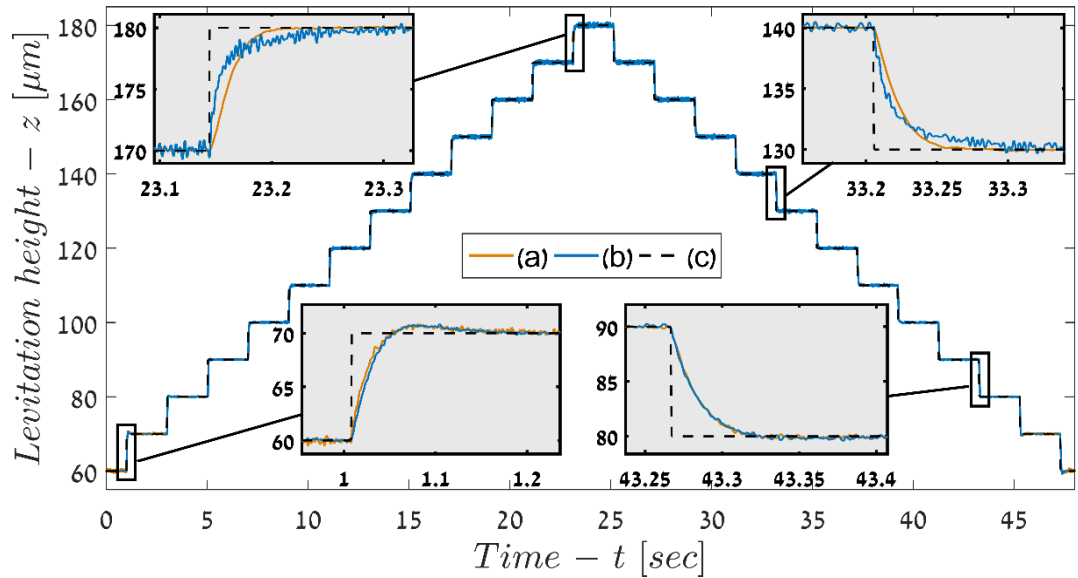

Fig. 9. Comparison between theoretical dynamic responses of the closed loop to the input reference (c), obtained (a) using a gain-scheduled controller, (b) using a linear controller designed for desired behavior around 80 µm.

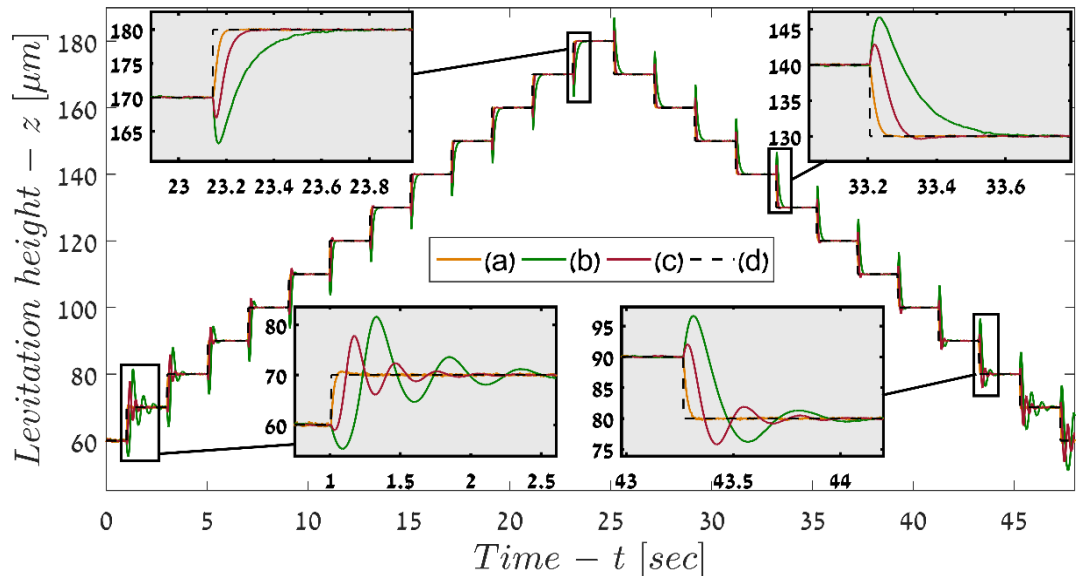

Fig. 10. Comparison between theoretical dynamic responses of the closed loop to the input reference (d), obtained (a) using a gain-scheduled controller, and using linear controllers designed for desired behavior around (b) 120 µm, (c) 160 µm.

### 4.2. Implementation of the control algorithm

The controller developed above was implemented experimentally, using the closed loop system illustrated schematically in Fig. 7. This system consists of the experimental setup displayed in Fig. 2, whose excitation frequency is determined by a Phase-locked loop resonance-tracking algorithm, an analog low-pass filter, reducing high frequency noise, and the controller designed above. The latter closes a feedback on the height of the levitated object, measured using a laser triangulation sensor (Keyence™, LK-H008). As shown in Fig. 7, the controller input is non-dimensional, thus the physical levitation height yielded by the sensor, was converted to a non-dimensional form using (2). Additionally, since the controller outputs the excitation magnitude $u$, but in practice it commands the amplitude of the input voltage $V$, the former was converted into the desired values using the relation found in the second preliminary experiment set (see sub-section 3.3).

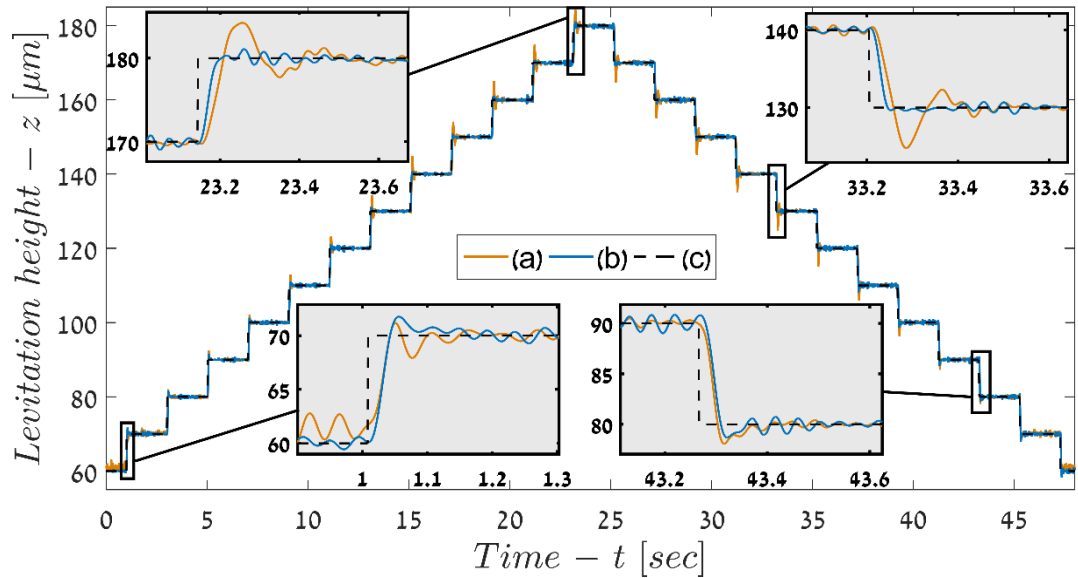

Fig. 11. Comparison between experimental dynamic responses of the closed loop to the input reference (c), obtained (a) using a gain-scheduled controller, (b) using a linear controller designed for desired behavior around 80 µm.

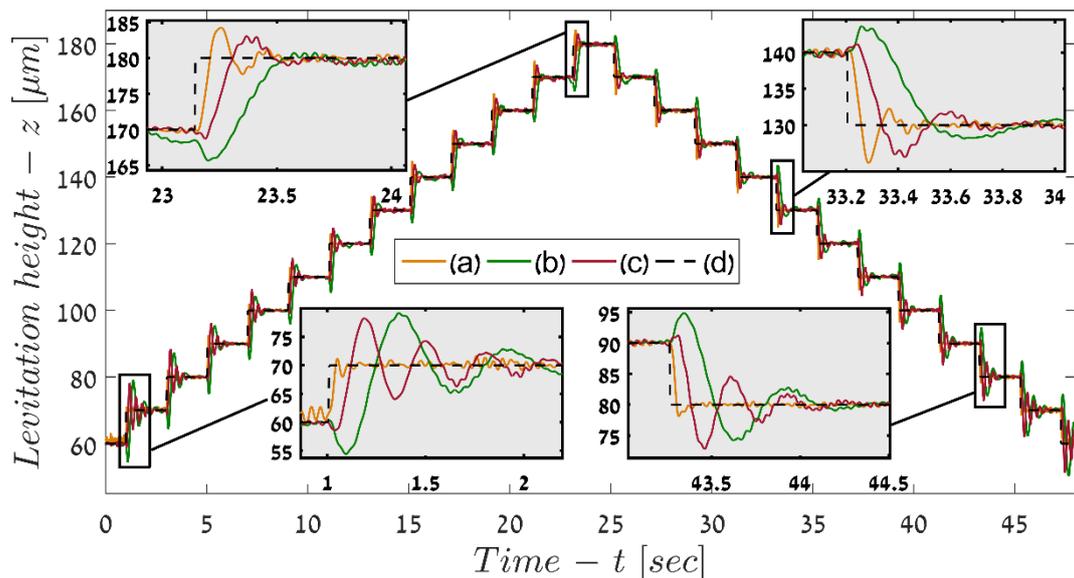

Fig. 12. Comparison between experimental dynamic responses of the closed loop to the input reference (d), obtained (a) using a gain-scheduled controller, and using linear controllers designed for desired behavior around (b) 120 µm, (c) 160 µm.

Fig. 11 and Fig. 12 present the experimental dynamic responses, corresponding to those discussed in the previous sub-section (Fig. 9-Fig. 10), filtered such that only frequencies lower than 35 Hz are taken into account. Namely, these figures display the dynamic responses of the closed loop system given in Fig. 7, achieved using the gain-scheduled controller (21)-(22), and using the three linear controllers designed similarly for a single design point. The parameters of the closed loop systems are given in Table 2, similarly to the simulations.

Fig. 12 shows that the superiority of the gain-scheduled controller, over the linear controllers designed around heights of 120 µm and 160 µm, is consistent with the simulations presented above. Moreover, as noticed in the simulations, the controllers designed around lower altitudes are sensitive to noise. Thus around 60 µm, the gain-scheduled controller did not keep the desired steady-state height. Yet, around 70 µm and beyond, all controllers tracked the commended steady-state height, accurately. However, as shown in Fig. 11, contrary to the prediction, in practice the linear controller designed around 80 µm yielded a better performance than the gain-scheduled controller. Moreover, the latter did not provide a critically damped behavior as expected. Nevertheless, it should be noted that the most significant deviations from the theory were obtained at significant altitudes, where the damping estimation is inadequate, leading to lower precision of the simplified model (13), on which the controller design is based.

## 5. Conclusions

It was shown that although high excitation frequencies are utilized when levitating planar objects by means of near-field acoustic levitation, the governing dynamics of the levitated objects is much slower. Consequently, to control the dynamics of near-field acoustically levitated objects, it is usually sufficient to consider merely their slow dynamics, implying that a relatively slow control system is adequate. However, due to the nonlinear behavior of the gas trapped between the driving surface and the levitated object, to achieve satisfactory performance, the control algorithm should be nonlinear as well.

In this paper, a gain-scheduled controller was designed based on a novel semi-analytical model, capturing the slow dynamics of a near-field acoustically levitated object, using a single, second order ordinary differential equation. The gain-scheduled controller exhibited adequately good performance, although the simplified model on which it is based underestimates the real damping of the system at significant air-gaps. Thus, it is the authors' opinion that based on the performance presented above, the simplified model developed in this paper is suitable to serve as the basis for more sophisticated model based controllers. Moreover, it is safe to assume that an empirically corrected damping term would have resulted in a superior controller.

Finally, although it is not the focus of this paper, the importance of the phase-locked loop, determining the excitation frequency, should be emphasized. Since the air-gap between the driving surface and the levitated object affects the stiffness and damping of the piezoelectric actuator (e.g. [7]), the resonance frequency of the actuator depend on the levitation height. Moreover, additional causes such as temperature and moisture variations, change the resonance frequency as well. Obviously, due to the low damping of the actuator, small deviations from the resonance, can reduce the excitation magnitude dramatically. Thus, working with a constant excitation frequency seems impractical, and so, a resonance-tracking algorithm must be an integral part of any control loop, governing the dynamics of near-field acoustically levitated objects.

## Acknowledgements

This research was funded by the Israeli Ministry of commerce under the Metro 450, Magnet program and the Ministry of Science, Technology and Space.

The authors would like to express their gratitude to Dr. Nadav Cohen who designed the experimental setup, and Mr. Ran Shaham who built the resonance-tracking loop and helped implementing the control algorithm.

Special thanks are to Nova Measuring Instruments for the cooperation, assistance and joint work on acoustic levitation of silicon wafers (www.novameasuring.com).

## References

[1]  G. Reinhart, J. Hoeppner, Non-contact handling using high-intensity ultrasonics, CIRP Ann. Technol. 49 (2000) 5–8.

[2]  A. Minikes, I. Bucher, Coupled dynamics of a squeeze-film levitated mass and a vibrating piezoelectric disc: numerical analysis and experimental study, J. Sound Vib. 263 (2003) 241–268.

[3]  Y. Wang, B. Wei, Mixed-Modal Disk Gas Squeeze Film Theoretical and Experimental Analysis, Int. J. Mod. Phys. B. 27 (2013).

[4]  D. Ilssar, I. Bucher, On the slow dynamics of near-field acoustically levitated objects under High excitation frequencies, J. Sound Vib. 354 (2015) 154–166.

[5]  D. Ilssar, I. Bucher, N. Cohen, Structural optimization for one dimensional acoustic levitation devices – Numerical and experimental study, in: ISMA 2014 Int. Conf. Noise Vib. Eng., Leuven Belgium, 2014.

[6]  W.A. Gross, L.A. Matsch, V. Castelli, A. Eshel, J.H. Vohr, M. Wildmann, Fluid film lubrication, John Wiley and Sons, Inc., New York, NY, 1980.

[7]  W.E. Langlois, Isothermal squeeze films, DTIC Document, 1961.

[8] M. Géradin, D. Rixen, Mechanical vibrations: theory and application to structural dynamics, Wiley New York, 1994.

[9] A. Minikes, I. Bucher, Comparing numerical and analytical solutions for squeeze-film levitation force, J. Fluids Struct. 22 (2006) 713–719.

[10] H. Nomura, T. Kamakura, K. Matsuda, Theoretical and experimental examination of near-field acoustic levitation, J. Acoust. Soc. Am. 111 (2002) 1578–1583.

[11] A. Minikes, I. Bucher, S. Haber, Levitation force induced by pressure radiation in gas squeeze films, J. Acoust. Soc. Am. 116 (2004) 217–226.

[12] E. Matsuo, Y. Koike, K. Nakamura, S. Ueha, Y. Hashimoto, Holding characteristics of planar objects suspended by near-field acoustic levitation, Ultrasonics. 38 (2000) 60–63.

[13] G.-C. Hsieh, J.C. Hung, Phase-locked loop techniques. A survey, Ind. Electron. IEEE Trans. 43 (1996) 609–615.

[14] J.-P. Yonnet, Permanent magnet bearings and couplings, Magn. IEEE Trans. 17 (1981) 1169–1173.

[15] K.J. Aström, R.M. Murray, Feedback systems: an introduction for scientists and engineers, Princeton university press, 2010.

[16] I. Kaminer, A.M. Pascoal, P.P. Khargonekar, E.E. Coleman, A velocity algorithm for the implementation of gain-scheduled controllers, Automatica. 31 (1995) 1185–1191. doi:10.1016/0005-1098(95)00026-S.

[17] W.J. Rugh, J.S. Shamma, Research on gain scheduling, Automatica. 36 (2000) 1401–1425.

[18] H.K. Khalil, Nonlinear systems, Prentice hall New Jersey, 2002.